\documentclass[11pt,a4paper]{article}
\usepackage[T1]{fontenc}
\usepackage[utf8]{inputenc}
\usepackage{lmodern}
\usepackage{amsmath,amssymb,bm}
\usepackage[margin=2.5cm]{geometry}
\usepackage{xcolor}
\usepackage{tikz}
\usetikzlibrary{arrows.meta,decorations.pathmorphing,patterns}
\usepackage[numbers,sort&compress]{natbib}
\usepackage{hyperref}
\hypersetup{colorlinks=true,linkcolor=blue!50!black,citecolor=blue!50!black,urlcolor=blue!50!black}

\newif\ifmarked
\markedfalse
\newcommand{\rev}[1]{\ifmarked{\color{red}#1}\else#1\fi}

\newcommand{\vect}[1]{\bm{#1}}
\newcommand{\Hk}{\mathrm{H}}

\title{Newton's second law: not so elementary\\(as it may seem)}
\author{Ledo Stefanini$^{1}$ \ and Giancarlo Reali$^{2}$\thanks{Corresponding author: reali@unipv.it}\\[6pt]
\small $^1$Accademia Nazionale Virgiliana, Via dell'Accademia 47, 46100 Mantova, Italy\\[3pt]
\small $^2$Dipartimento di Ingegneria Industriale e dell'Informazione, Universit\`a di Pavia,\\
\small via Ferrata 5a, 27100 Pavia, Italy}
\date{}

\begin{document}
\maketitle

\begin{abstract}
\rev{Introductory courses present Newton's second law either as a definition of force or as a
law of nature, and in both cases leave unexplained why it holds only in a special class of
reference frames. We give an operational reading of the law built on a single instrument, the
spring balance. Its reading is a local fact on which all observers agree; in a given laboratory it
defines the masses of bodies and a vector $\vect{\gamma}$ characterising the laboratory.
Laboratories with $\vect{\gamma}=0$ are, by definition, inertial. The empirical content of the
second law is then the statement $\vect{\gamma}=-K\vect{A}$, where $\vect{A}$ is the acceleration
of the laboratory measured kinematically from any inertial laboratory, and $K$ is a universal
constant that the choice of the newton as unit of force makes invisible. In this reading a spring
balance is an accelerometer and an accelerometer is a dynamometer: both measure the specific
force $\vect{a}-\vect{g}$ familiar from inertial navigation. We discuss the
hypotheses involved (universality of free fall, absence of rotation, locality) and the relation
with Newton's own formulation.}
\end{abstract}

\section{Introduction}

The adjective `Newtonian' is used conventionally for classical mechanics. Yet over three
centuries the reflection on mechanical phenomena has led to a deep revision of the principles on
which Newton built his theory, and has taught us to recognise the dangers hidden in the
anthropomorphic conceptions that form the intuitive basis of introductory physics. One might
expect the critical work of Mach~\cite{Mach} and Poincar\'e~\cite{Poincare} to have had some
impact on the teaching of elementary mechanics, which, while still called `Newtonian', should not
be identified with that of the \emph{Principia}. More than a century later, this is not the case:
almost all introductory textbooks follow the historical path in their exposition of dynamics.
The reasons for this conservatism are many.

Treatises of analytical mechanics generally follow the school of Kirchhoff~\cite{Kirchhoff} and
Sommerfeld~\cite{Sommerfeld}, for which the second law is nothing but a definition of force.
Another school holds that the principles of Newtonian mechanics are not definitions but genuine
laws of nature; a well-known representative of this position is Feynman, who devoted the twelfth
chapter of his \emph{Lectures} to a sharp and witty discussion of the second
law~\cite{Feynman}. \rev{The question has been debated repeatedly in the didactic
literature~\cite{Eisenbud,Jammer,Wilczek}.}

In this work we propose a reading suggested by the principle of relativity.
\rev{The reading of a dynamometer is a local event (the extension of a spring) on which all
observers agree. Acceleration, on the contrary, is a kinematic quantity that depends on the
reference frame, and is common only to the frames of a special class, the inertial ones. Since the
second law connects the two, it can hold only in that class.} Moreover, it is not easy to
separate the measurement of a force from that of an acceleration. It would be more appropriate to
say that accelerometers measure forces while dynamometers measure accelerations. \rev{The aim
of this paper is to give this sentence a precise meaning (Sec.~\ref{sec:formal}).}

\section{Is acceleration an observable?}

The formal definition of acceleration as the second derivative of position with respect to time
has no direct operational meaning: the concept is meaningful only as the limit of an average
acceleration. No instrument measures the instantaneous acceleration of an object directly; to
measure it we must connect it with another observable quantity.

For example, we may track an approaching object by radar, sending it a wavepacket of known
central frequency $\nu_0$ and measuring the frequency shift of the echo.
\rev{To first order in $v/c$ the shift is proportional to the radial velocity,
$\Delta\nu \simeq 2\nu_0 v_r/c$, so that the radial acceleration is obtained from its rate of
change, $a_r \simeq (c/2\nu_0)\,\mathrm{d}\Delta\nu/\mathrm{d}t$; the full vector requires
tracking from several stations.} This measurement, however, involves only instruments of our
laboratory and gives a \emph{relative} acceleration: it does not tell whether the body or the
observer is accelerating.

We may instead place an accelerometer on the body, as on board Formula One cars. The instrument
transmits its data to the observer, and it is immediately clear whether the observed body or the
observer is accelerating. Yet even an accelerometer does not measure acceleration directly: it
measures the deformation of an elastic element, which is assumed to be proportional to the
acceleration. \rev{We shall see that this assumption is the second law itself.}

\section{Spring balances in Galileo's cabin}

Galileo opens his celebrated argument for the relativity of uniform motion with an
invitation~\cite{Galileo}:
\begin{quote}
`Shut yourself up with some friend in the main cabin below decks on some large ship, and have
with you there some flies, butterflies, and other small flying animals\ldots'
\end{quote}
\rev{Flies and butterflies cannot reveal whether the ship is at rest or in uniform motion.
Let us add to the equipment of the cabin a more modest instrument, which will instead reveal
whether the ship is accelerating.} It is a spring balance: a spring with two hooks, one to anchor
it to a support, the other to hang bodies on. Its obvious use is to attribute to a body a property
called weight, and the physicists in the cabin will use it in this way. Note however that what the
inhabitants of the cabin call a measurement of weight is, for us observing the ship from outside, a
measurement of acceleration.

Problems arise when the physicists in the cabin compare their results with those of other
laboratories. If these are at different heights or latitudes, or inside elevators or spaceships,
the same body will be found to have different weights. We are thus led to attribute to the body a
new quantity, its mass $m$, and to the laboratory a quantity $\vect{\gamma}$, such that
\begin{equation}
\vect{w} = m\,\vect{\gamma}.
\label{eq:w}
\end{equation}
Here $\vect{w}$ is the weight read on the balance, $m$ depends only on the body and
$\vect{\gamma}$ only on the laboratory; we call $\vect{\gamma}$ the gravity of the laboratory.

\rev{Equation~\eqref{eq:w} is not a convention: it contains the empirical fact that the ratio of
the weights of two bodies is the same in every laboratory. This is the universality of free fall,
or equality of inertial and gravitational mass, tested by E\"otv\"os and today verified to about
one part in $10^{15}$~\cite{Microscope}.} It is a far from trivial hypothesis that a sample of
mass can be shared among laboratories (meaning two bodies that have the same weight in one
laboratory). With such samples the balances can be calibrated and used as gravimeters,
instruments that characterise not single bodies but laboratories.

If weights are measured in a unit defined by the extension of a standard spring, which we call
the Hooke ($\Hk$), and masses in kilograms, $\vect{\gamma}$ is measured in $\Hk/\mathrm{kg}$.
If we want a mechanics shared among all laboratories, each of them has to measure its own
$\vect{\gamma}$. Among all laboratories, some have a singular property: their gravimeters always
give a null result. For these laboratories the construction of a shared mechanics is particularly
simple. We call laboratories with $\vect{\gamma}=0$ \emph{inertial}; examples are artificial
satellites and all vehicles in free fall.

\section{Spring balances as accelerometers}

To single out the laboratories that deserve the name inertial, a spring balance is enough. If the
physicists of two inertial laboratories also use radar, they find a further property: each sees
the other in uniform rectilinear motion. Inertial laboratories are privileged because the same
simple laws of mechanics hold in all of them; these laws do not apply in the other laboratories,
which have $\vect{\gamma}\neq 0$.

If from an inertial laboratory we observe a non-inertial one by radar, we find that it is
accelerating with respect to us. Its physicists, who already know they are in a non-inertial
laboratory because their balances do not read zero, find in turn that we are accelerating with
respect to them. Kinematically the situation is symmetric: if A accelerates with respect to B,
then B accelerates with respect to A. Dynamically it is not: the balances tell which of the two
is inertial, and the shared laws of the inertial class entitle us to say that the non-inertial
laboratory is the one that accelerates. \rev{This is the content of the principle of equivalence
in the form given by Einstein in 1907~\cite{Einstein1907}: a laboratory at rest in a gravitational
field cannot be distinguished, by local experiments, from a laboratory accelerating in the
opposite direction with respect to an inertial one. Its validity is local, as discussed in
Sec.~\ref{sec:limits}.}

Accepting this principle changes the meaning of the readings of the balance. For the physicists
inside a non-inertial laboratory the balance signals the weight of a body, and the proportionality
of weight to mass is no surprise: the concept of mass arises precisely from the requirement of a
quantity that characterises a single body and is invariant among laboratories, as expressed by
\rev{Eq.~\eqref{eq:w}}. For the physicists of an inertial laboratory, what the balance measures in
the non-inertial laboratory is not gravity but the acceleration of the laboratory itself: the
balance acts not as a gravimeter but as an accelerometer. Every measurement of acceleration made
by inertial observers corresponds to a measurement of force made by observers in the accelerated
laboratory. The deeper meaning of Newton's second law is the proportionality between the gravity
measured in a laboratory and the acceleration of that laboratory measured from an inertial one:
\begin{equation}
\rev{\vect{\gamma} = -K\,\vect{A},}
\label{eq:K}
\end{equation}
where $\vect{\gamma}$ is measured with a gravimeter and $\vect{A}$ by kinematic methods.
\rev{The minus sign records that a laboratory accelerating upwards weighs bodies downwards
(Fig.~\ref{fig:labs}).} The constant $K$ has dimensions
\begin{equation}
[K] = \rev{\frac{\Hk/\mathrm{kg}}{\mathrm{m/s^2}} = \frac{\Hk\,\mathrm{s^2}}{\mathrm{kg\,m}}.}
\label{eq:dimK}
\end{equation}

\begin{figure}[t]
\centering
\begin{tikzpicture}[>={Straight Barb[length=2.2mm]},thick,scale=0.95]
  \newcommand{\lab}[3]{%
    \begin{scope}[shift={(#1,0)}]
      \draw (0,0) rectangle (2.4,3.2);
      \draw[line width=1.2pt] (0.7,3.2) -- (1.7,3.2);
      \draw[decorate,decoration={coil,aspect=0.4,segment length=2.5pt,amplitude=3pt}]
            (1.2,3.2) -- (1.2,#2);
      \fill[gray!40] (0.95,#2) rectangle (1.45,{#2-0.5});
      \draw (0.95,#2) rectangle (1.45,{#2-0.5});
      \node at (1.2,-0.45) {#3};
    \end{scope}}
  \lab{0}{1.6}{(a) at rest on Earth}
  \fill[pattern=north east lines] (-0.3,-0.15) rectangle (2.7,0);
  \draw[->,red!70!black] (2.0,2.0) -- (2.0,1.0) node[right] {$\vect{\gamma}$};
  \draw[->,blue!60!black] (-0.35,1.0) -- (-0.35,2.2) node[left] {$\vect{A}$};
  \lab{4.2}{2.3}{(b) free fall}
  \node at (5.4,0.8) {$\vect{\gamma}=0$};
  \lab{8.4}{1.6}{(c) rocket, deep space}
  \draw[->,red!70!black] (10.4,2.0) -- (10.4,1.0) node[right] {$\vect{\gamma}$};
  \draw[->,blue!60!black] (8.05,1.0) -- (8.05,2.2) node[left] {$\vect{A}$};
\end{tikzpicture}
\caption{\rev{The same spring balance in three laboratories. In (a) and (c) the balance gives the
same reading: the gravity $\vect{\gamma}$ of the laboratory points opposite to its acceleration
$\vect{A}$ with respect to the inertial class, Eq.~\eqref{eq:K}. In (b) the laboratory is
inertial and the balance reads zero.}}
\label{fig:labs}
\end{figure}

Like all universal constants~\cite{LevyLeblond}, $K$ plays the role of a `conceptual
synthesiser': it expresses the unification of two distinct physical concepts into a single one of
more general validity. This is not new in physics. The mechanical equivalent of heat $J$ is
another example, and so is Boltzmann's constant $k$, which synthesises temperature and kinetic
energy.

\section{A formal statement}
\label{sec:formal}

\rev{We now collect the argument in a compact form. We consider laboratories that are small and
non-rotating (the second condition can be checked with gyroscopes), and we use three
empirical laws.}

\rev{\emph{(E1) Mass and gravity.} For every body held at rest by a spring balance in a laboratory
$L$, the reading is $\vect{w}=m\,\vect{\gamma}(L)$, with $m$ depending only on the body and
$\vect{\gamma}(L)$ only on the laboratory. Laboratories with $\vect{\gamma}=0$ are called
inertial.}

\rev{\emph{(E2) The inertial class.} Two inertial laboratories move uniformly with respect to each
other. Hence the acceleration $\vect{A}(L)$ of a laboratory $L$, measured by kinematic means, is
the same for every inertial observer: it is a property of $L$ relative to the inertial class, not
to a particular observer.}

\rev{\emph{(E3) Second law.} For every laboratory, $\vect{\gamma}(L)=-K\vect{A}(L)$, with $K$ a
universal constant.}

\rev{To these we add the static rule that readings of balances acting on the same body compose
as vectors (parallelogram of forces). Consider now a body of mass $m$ held at rest in $L$ by a
spring. The force exerted by the spring on the body is opposite to the weight,
$\vect{F}=-\vect{w}$, and the body shares the acceleration of the laboratory, $\vect{a}=\vect{A}$.
From (E1) and (E3),
\begin{equation}
\vect{F} = -m\,\vect{\gamma} = K\,m\,\vect{a}.
\label{eq:Fma}
\end{equation}
A body that moves in $L$ is at rest in its own non-rotating comoving laboratory, to which the same
argument applies; Eq.~\eqref{eq:Fma} therefore holds for any body, with $\vect{F}$ the resultant of
the forces read on springs (and, more generally, on any contact interaction calibrated against
them) and $\vect{a}$ its acceleration with respect to the inertial class. This is Newton's second
law with an explicit proportionality constant. The choice $K=1$ defines the newton as unit of
force, $1\,\mathrm{N}=1\,\mathrm{kg\,m/s^2}$.}

\rev{\emph{Newtonian translation.} Newton refers accelerations to a single frame $N$ (the `fixed
stars') and describes gravitation as a force $m\vect{g}_N(\vect{r})$. In this language a
freely falling laboratory accelerates with $\vect{g}_N$, so that the acceleration of $L$ with
respect to the local inertial laboratory is $\vect{A}=\vect{a}_N-\vect{g}_N$. With $K=1$,
Eqs.~\eqref{eq:K} and \eqref{eq:Fma} become
\begin{equation}
\vect{f} \equiv \frac{\vect{F}}{m} = -\vect{\gamma} = \vect{a}_N-\vect{g}_N,
\qquad\text{i.e.}\qquad
m\,\vect{a}_N = \vect{F} + m\,\vect{g}_N .
\label{eq:f}
\end{equation}
The two descriptions contain the same physics and differ only in where gravitation is placed: as
a force on the right-hand side (Newton) or in the definition of the inertial class (here). For a
laboratory at rest on the ground, $\vect{a}_N\simeq 0$ and $\vect{\gamma}=\vect{g}_N$, the usual
weight per unit mass. The quantity $\vect{f}$ is what inertial-navigation engineers call the
\emph{specific force}~\cite{Titterton}: it is the only quantity an accelerometer can sense, and
the accelerometer cannot separate its two terms. In this precise sense an accelerometer measures
a force (per unit mass), and a dynamometer, through $\vect{\gamma}=-K\vect{A}$, measures an
acceleration.}

\section{Limits of validity}
\label{sec:limits}

\rev{The laws (E1)--(E3) hold under conditions worth stating explicitly.}

\rev{\emph{Locality.} In a real gravitational field $\vect{\gamma}$ is not uniform. Across a
freely falling laboratory of size $d$ at distance $R$ from the centre of the Earth, residual
(tidal) gravities of order $g\,d/R$ remain; for $d=10$~m this is about $10^{-5}\,g$. A laboratory
is inertial only to the extent that these residues are below the sensitivity of its balances.
This is why inertial laboratories are defined locally, and why no global inertial frame exists in
general relativity.}

\rev{\emph{Rotation.} In a rotating laboratory a body at rest weighs differently at different
places (centrifugal effect), and a moving body is deflected (Coriolis effect). A single balance
cannot detect a pure rotation about its own position, so the non-rotating condition must be
established separately, with gyroscopes or by observing distant objects.}

\rev{\emph{Universality of free fall.} If two bodies of different composition fell with different
accelerations, the ratio of their weights would depend on the laboratory, (E1) would fail, and no
laboratory could have $\vect{\gamma}=0$ for all bodies at once. The whole construction rests on
this experimental fact~\cite{Microscope}.}

\section{The fate of universal constants}

This is the right moment to re-read Newton's formulation~\cite{Newton}:
\begin{quote}
\emph{Mutationem motus proportionalem esse vi motrici impressae, \& fieri secundum lineam
rectam qua vis illa imprimitur.}
\end{quote}
In the current formulation the adjective \emph{proportionalem} is cheerfully replaced by
\emph{aequalem}: the constant of proportionality has been swept under the rug. This should not
let us forget that the unit of force was chosen precisely for this purpose: the newton is defined
so that the constant between the gravity of a laboratory and its acceleration disappears. Such is
the fate of universal constants. As their role of conceptual synthesisers becomes part of the
common background of physics, they are reduced to conversion factors between units and finally,
with a suitable choice of units, they disappear. This is what happened to the mechanical
equivalent of heat $J$ and, in many respects, to the speed of light $c$.

\section{Conclusions}

\rev{Newton's second law is neither a mere definition nor a law that can be stated without
specifying the frames in which it holds. Read operationally, it is the empirical identity
$\vect{\gamma}=-K\vect{A}$ between two independently measured quantities: the gravity of a
laboratory, read on a spring balance inside it, and its acceleration, measured kinematically from
outside by any inertial observer. The definition of mass rests on the universality of free fall,
the definition of the inertial class on the null reading of balances, and the constant $K$,
hidden by the choice of the newton, records the unification of the two concepts. We believe that
this reading, which requires nothing beyond a spring balance and the idea of a freely falling
laboratory, can be presented in introductory courses, where it also prepares the ground for the
principle of equivalence.}

\end{document}